\documentclass[%
 aip,
 amsmath,amssymb,
 reprint,%
]{revtex4-1}

\usepackage{graphicx}
\usepackage{dcolumn}
\usepackage{bm}

\usepackage[utf8]{inputenc}
\usepackage[T1]{fontenc}
\usepackage{mathptmx}
\usepackage{etoolbox}
\usepackage{upgreek}
\usepackage{multirow}
\usepackage{hyperref}
\usepackage{xcolor}
\usepackage{colortbl}

\makeatletter
\def\@email#1#2{%
 \endgroup
 \patchcmd{\titleblock@produce}
  {\frontmatter@RRAPformat}
  {\frontmatter@RRAPformat{\produce@RRAP{Author to whom correspondence may be addressed: \href{mailto:#2}{#2}}}\frontmatter@RRAPformat}
  {}{}
}%
\makeatother
\begin{document}

\preprint{AIP/123-QED}

\title{Fully-digital platform for local ultra-stable optical frequency distribution\\}
\author{Martina Matusko}
 \affiliation{Université de Franche-Comté, SUPMICROTECH, CNRS, institut FEMTO-ST, F-25000 Besançon, France}
\author{Ivan Ryger}
 \affiliation{Université de Franche-Comté, SUPMICROTECH, CNRS, institut FEMTO-ST, F-25000 Besançon, France}
\author{Gwenhaël Goavec-Merou}
 \affiliation{Université de Franche-Comté, SUPMICROTECH, CNRS, institut FEMTO-ST, F-25000 Besançon, France}
\author{Jacques Millo}
 \affiliation{Université de Franche-Comté, SUPMICROTECH, CNRS, institut FEMTO-ST, F-25000 Besançon, France}
\author{Clément Lacroûte}
 \affiliation{Université de Franche-Comté, SUPMICROTECH, CNRS, institut FEMTO-ST, F-25000 Besançon, France}
\author{\'Emile Carry}
 \affiliation{Université de Franche-Comté, SUPMICROTECH, CNRS, institut FEMTO-ST, F-25000 Besançon, France}
\author{Jean-Michel Friedt}
 \affiliation{Université de Franche-Comté, SUPMICROTECH, CNRS, institut FEMTO-ST, F-25000 Besançon, France}
\author{Marion Delehaye}
 \affiliation{Université de Franche-Comté, SUPMICROTECH, CNRS, institut FEMTO-ST, F-25000 Besançon, France}
 \email{marion.delehaye@femto-st.fr}

\begin{abstract}
This article reports on the use of an FPGA platform for local ultra-stable optical frequency distribution through a 90~m-long fiber network. This platform is used to implement a fully digital treatment of the Doppler-cancellation scheme required by fiber links to be able to distribute ultra-stable frequencies. We present a novel protocol that uses aliased images of a digital synthesizer output to directly generate signals above the Nyquist frequency. This approach significantly simplifies the set-up, making it easy to duplicate within a local fiber network. We demonstrate performances enabling the distribution of an optical signal with an instability below $10^{-17}$ at one second at the receiver end. We also use the board to implement an original characterization method. It leads to an efficient characterization of the disturbance rejection of the system, that can be realized without accessing the remote output of the fiber link.
\end{abstract}

\maketitle

\section{\label{Introduction}Introduction}
Optical atomic oscillators can reach the $10^{-19}$ level of instability after $10^5$~s averaging time~\cite{McGrew:2018}. Together with optical clocks uncertainties that can be below the $10^{-18}$ level~\cite{Brewer2019}, this paves the way to numerous applications, ranging from new GNSS architecture~\cite{Schuldt:2021} or improved geodesy~\cite{Takamoto2021} to advances in fundamental science, such as the measurement of the relativistic redshift~\cite{Pavlis_2003, Chou2010}, or searches for dark matter~\cite{Safronova:2018}. 

Fibers links have proven to be the most robust~\cite{narbonneau:2006, cantin:2021} and efficient way to transfer ultra-stable signals over distances up to thousands of kilometers~\cite{Ma1994,  Predehl:04, Newbury:07,  Grosche:09, Fujieda:11, Lopez:12, Calonico2014, Lisdat:2015,   Husmann2021}. Long-distance fiber networks now connect many European laboratories using dedicated infrastructures~\cite{Lisdat:2015, Roberts_2020}. 
Here, we focus on stable frequency distribution within an institute, where several experiments distant by typically tens to hundred meters may require access to ultra-stable optical signals.
Fractional frequency instabilities that can be reached in research facilities range from $10^{-15}$~ to $10^{-17}$ at one second~\cite{refimeve, Oelker2019}. This sets the targeted frequency instability of the local distribution fiber network discussed in this paper.
Fibers can guide an optical signal over distances with limited power loss, but the purity of the signals can be degraded while propagating because of thermal and mechanical perturbations affecting the optical length of the fiber.
Phase noise induced in fiber links is usually tracked and compensated for by Doppler cancellation techniques~\cite{Williams:08, Calosso:15} in a phase-locked loop (PLL). These traditionally relied on analog electronics for tracking and correcting the induced phase noise. Powerful alternatives to analog devices are digital electronics. They are very suited to ultra-stable frequency dissemination within an institute, as they are less sensitive to environmental perturbations and possess the advantages of versatility, reconfigurability, and easy duplication~\cite{Ning:2013, Baldwin:12, Cardenas:2016, Plante2018}.

In this paper, we demonstrate a fully-digital and open-source Doppler cancellation platform based on a Red Pitaya $\mathrm{SDR^{lab}} 122-16$ (RP16) board for a 90~m fiber link at 1542~nm (194~THz) for frequency dissemination within the FEMTO-ST institute. Longer links can also be compensated using this setup, at the price of a reduced lock bandwidth. We make use of the capabilities of the open-source FPGA platform \emph{oscimpDigital}, available on github under the CeCILL licence~\cite{oscimp}.
First, we assess the performances of the newly available RP16 and compare them with the previous Red Pitaya  $\mathrm{STEM^{lab} 125-14}$ (RP14). We then explore the effects of a fully-digital signal treatment using the analog-to-digital (ADC) and digital-to-analog (DAC) converters of RP16. In particular, we make a novel use of aliasing at the output of a direct digital synthesizer (DDS) and characterize the delays due to different blocks of the set-up and their contribution to the PLL bandwidth.
Lastly, we present an automatic method to measure the disturbance rejection of the PLL in order to reach the best noise-cancellation performances.

\section{\label{digital_platforms}Digital platforms}
\begin{figure}[h!]
    \centering
    \includegraphics[width=0.6\columnwidth]{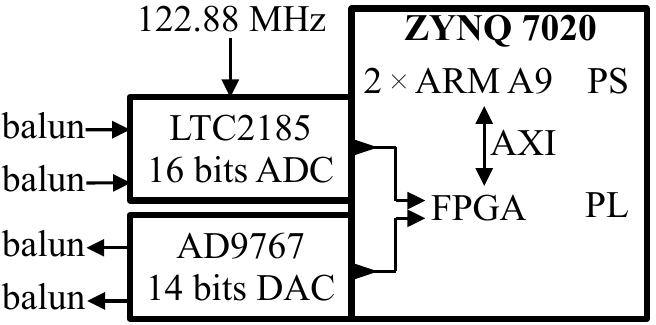}
    \caption{Schematic description of the relevant blocks of RP16. }
    \label{fig:RP_schematics}
\end{figure}

Ultra-stable frequency dissemination through fiber links has already been performed using an RP14 board and showed promising results \cite{9599700, Cardenas:2016, 8856059}. An RP16 board recently became available with a more powerful FPGA to embed more logical functions and increase processing capabilities. 
It is explicitly designed for radio frequency instrumentation, with an analog front-end design targeted towards using higher Nyquist zones and a readily accessible external reference input port. A schematic view of the constitutive blocks of RP16 is shown Fig.~\ref{fig:RP_schematics}. Extra key characteristics of RP16 and RP14 are indicated in Table~\ref{RP_comparison}, as well as critical resource use for the schemes used in this article. 

\begin{table}[h!]
\begin{tabular}{c|c|c|}
\cline{2-3}
\multicolumn{1}{l|}{}                                                                               & \textbf{RP14}     & \textbf{RP16}     \\ \hline
\multicolumn{1}{|c|}{\textbf{ADC}}                                                                  & 14 bits           & 16 bits           \\ \hline
\multicolumn{1}{|c|}{\textbf{\begin{tabular}[c]{@{}c@{}}ADC effective \\  number of bits\end{tabular}}}                                                             & 11.8                & 12.5              \\ \hline
\multicolumn{1}{|c|}{\textbf{\begin{tabular}[c]{@{}c@{}}Measurement \\ bandwidth\end{tabular}}} & 52 MHz\footnote{The measurement bandwidth can be increased to reach the ADC track-and-hold bandwidth of 750~MHz by removing the anti-aliasing filter before the ADC.}           & 550 MHz           \\ \hline
\multicolumn{1}{|c|}{\textbf{DAC}}                                                                  & 14 bits           & 14 bits           \\ \hline
\multicolumn{1}{|c|}{\textbf{FPGA}}                                                                 & Xilinx Zynq 7010  & Xilinx  Zynq 7020 \\ \hline
\multicolumn{1}{|c|}{\textbf{Resource use}}                                                                 &   &  \\  \arrayrulecolor{lightgray}\hline    \arrayrulecolor{black}
\multicolumn{1}{|c|}{Dual PLL~\cite{9599700}}                                                                 &  75~\% of DSP\footnote{Digital Signal Processing} & 27~\% of DSP \\     \arrayrulecolor{lightgray}\hline    \arrayrulecolor{black}
\multicolumn{1}{|c|}{\begin{tabular}[c]{@{}c@{}}PLL with \\ characterization \\ (see Fig.~\ref{fig:double_iq_pid_vco_charac})\end{tabular}}                                                            & 94~\% of BRAM\footnote{Blocks of Random Access Memory}  & 44~\% of BRAM \\ \hline
\multicolumn{1}{|c|}{\textbf{\begin{tabular}[c]{@{}c@{}}Quartz crystal \\ (TCXO)\end{tabular}}}                                                       & FXO HC735-125 MHZ & ABLNO-122.88MHz   \\ \hline
\multicolumn{1}{|c|}{\textbf{Clock frequency}}                                                      & 125 MHz           & 122.88 MHz        \\ \hline

\end{tabular}
\vspace{+0.6 mm}
\caption{Key characteristics of RP14 and RP16 boards\\ }
\label{RP_comparison}
\end{table}

\begin{figure}[h!]
    \centering
    \includegraphics[width=0.95\columnwidth]{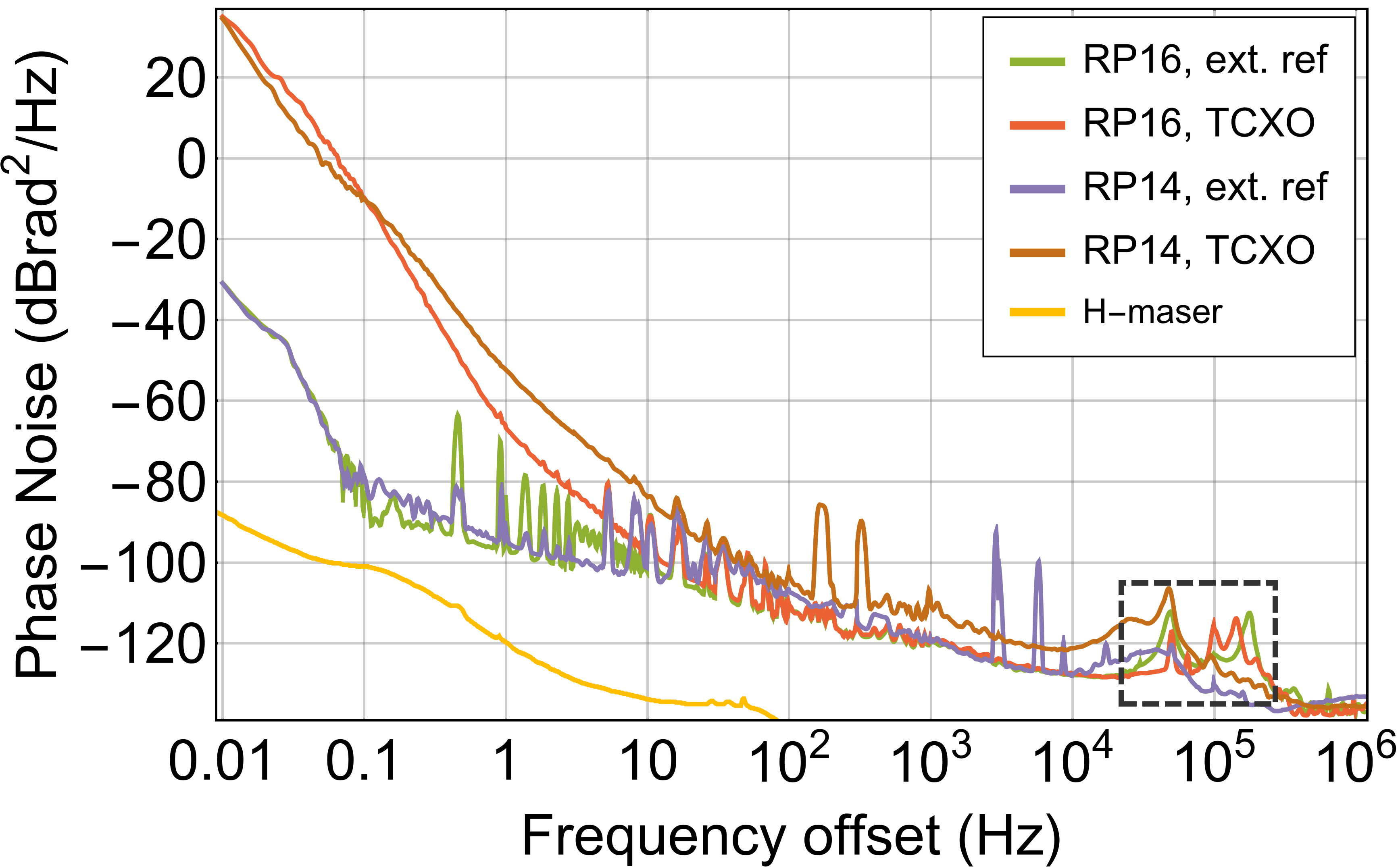}
    \caption{Phase-noise of a 10~MHz signal generated by Red Pitaya boards when clocked by their internal TCXO oscillator or by an external reference derived from an active hydrogen maser. Green: RP16 with external reference. Red: RP 16 with internal TCXO. Purple: RP14 with external reference. Brown: RP14 with internal TCXO. Yellow: hydrogen maser. The squared area indicates the spurious signals coming from the DAC voltage regulators.}
    \label{fig:out10}
\end{figure}

To assess these boards for metrology purposes, we first measured their output phase noise at 10~MHz with either their internal (temperature-compensated crystal oscillator -- TCXO) or an external (AD9915 DDS clocked by an active hydrogen maser) clock reference. 
The results are shown in Fig.~\ref{fig:out10}. 
In agreement with the specified "low-noise" characteristics of RP16 TCXO, its phase noise shows a 15~dB improvement close to 1~Hz with respect to RP14. However, below 0.1~Hz, the phase noise of $-10~\mathrm{dBrad^2/Hz}$ is similar for the two boards. 
Such performances of the boards in a Doppler-cancellation scheme for fiber links limit the relative instability of the frequency transfer to the  $10^{-15}$ range~\cite{Rubiola2022}, similar to the degradation of the signal stability we measured when propagating through 90~m-long optical fibers in a noisy environment. Clocking the boards with a stable external reference is therefore mandatory for most metrological applications. 
It should be noted that both the RP14 and the RP16 present spurious signals at 80 and 170~kHz with an amplitude of about $-91~\mathrm{dBrad^2/Hz}$ (see squared area Fig.~\ref{fig:out10}). These are generated by the DAC voltage regulators and cannot be removed easily\footnote{A low-noise version of Red Pitaya STEMlab 124-14 very recently became available, that claims improved linear voltage regulators.}. 
When the boards are used within a PLL, they may induce oscillations at these frequencies.

Regarding the physical board, the RP14 ADC has a low-pass, anti-aliasing filter that is not present on the RP16. This makes possible for the RP16 direct undersampled detection of signals at a frequency higher than $f_{\mathrm{clock}}/2$ (where $f_{\mathrm{clock}}$ is the clock frequency, see Table~\ref{RP_comparison}): an incoming signal at $f_{\mathrm{in}}>f_{\mathrm{clock}}/2$ is detected at a frequency $f'_{\mathrm{in}} = | f_{\mathrm{in}} - n\times f_{\mathrm{clock}}|$, where $n$ is an integer such that $0<f'_{\mathrm{in}}<f_{\mathrm{clock}}/2$.

Undersampling has already been used in the context of fiber links~\cite{9599700} with an RP14 that had to undergo ADC input circuit modification, while the use of an RP16 leads to a much simpler implementation.

\section{\label{dig_cancellation_link}Fully-digital cancellation of fiber links noise}

\begin{figure}[h!]
    \centering
    \includegraphics[width=0.95\columnwidth,
    clip]{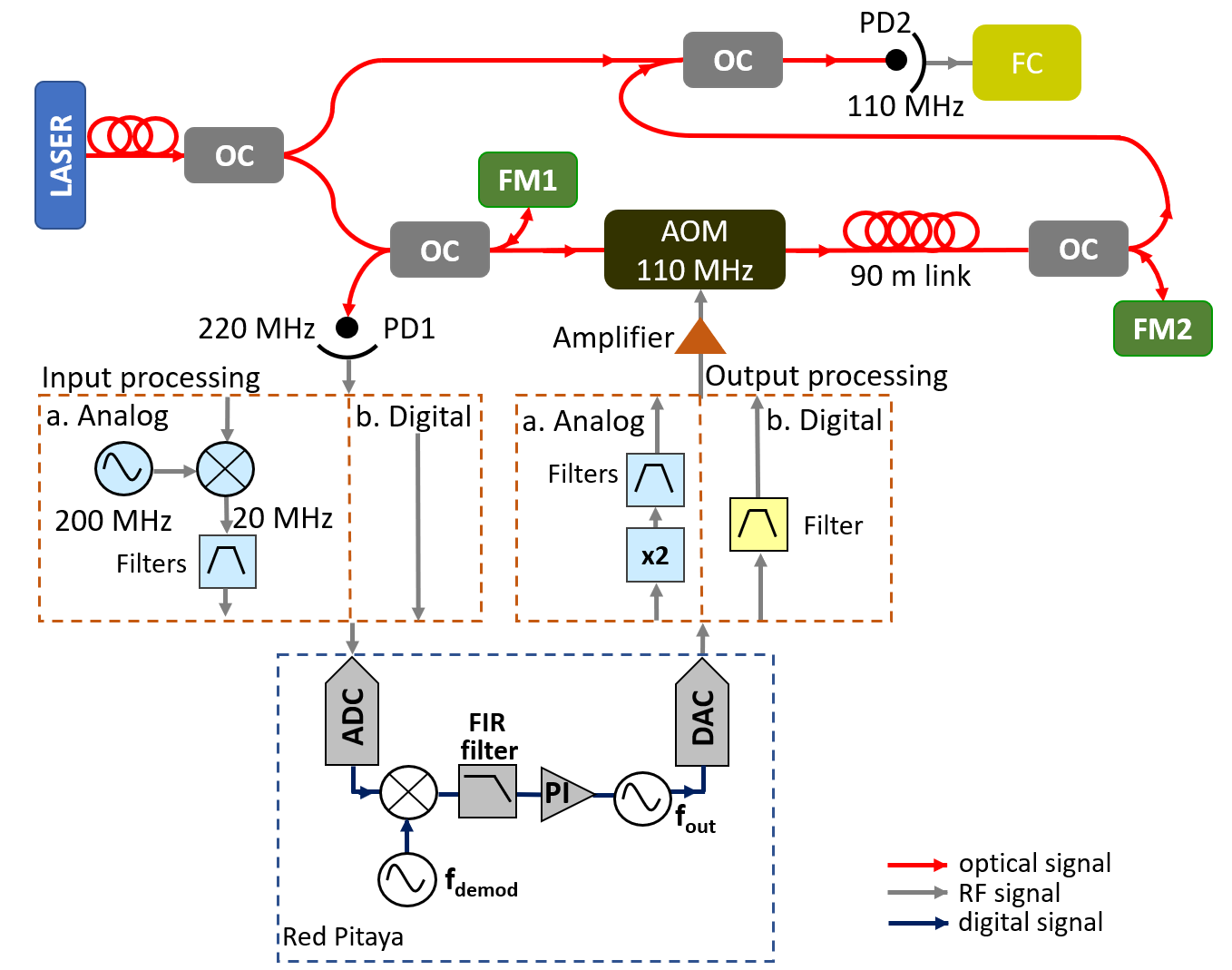}
    \caption{Doppler-cancellation set-up. OC: optical coupler; FM: Faraday mirror; PD: photodiode; FC: frequency counter; input and output processing: see text. PD1 detects the 220 MHz beatnote resulting from the return trip through the AOM. PD2 detects a beatnote at 110 MHz that is used to characterize the system.}
    \label{fig:exp_set-up}
\end{figure}

We performed Doppler cancellation of the phase noise arising from a 90~m-long fiber due to thermal and mechanical fluctuations. 
We first measured the phase noise of the system with and without the 90~m-long fiber and found that the presence of the fiber leads to a noise increase in the $1~\mathrm{Hz} - 10~\mathrm{kHz}$ range by up to 40~dB (at 1~Hz).
Our Doppler-cancellation set-up is based on a heterodyne Michelson interferometer as depicted in Fig.~\ref{fig:exp_set-up} and signal treatment is done with RP16. It is fully-fibered with single-mode non polarization-maintaining fibers.
One arm of the interferometer is a short fiber ended by a Faraday mirror (FM1) and plays the role of a reference arm. The second arm is composed of a fiber-coupled acousto-optic modulator (AOM\footnote{Gooch \& Housego T-M110-0.2C2J-3-F2S}) driven at $f_{\mathrm{AOM}} = 110~\mathrm{MHz}$ followed by the fiber link and a second fiber-coupled Faraday mirror (FM2). We measure the noise arising from the fiber link by generating a beatnote between the signal reflected by FM1 and the signal reflected by FM2 that is twice frequency shifted by the AOM at $f_{\mathrm{beatnote}} = 2 \times f_{\mathrm{AOM}}=220~\mathrm{MHz}$. The resulting beatnote is then detected on a photodiode (PD1) for the aim of the external phase disturbance correction.
For characterization purposes, the 90~m-long fiber does a round-trip, with the output of the link located within a few centimeters from its input. Part of the link output interferes with the signal coming directly from the master laser, and the resulting beatnote at 110~MHz is detected by a second photodiode (PD2) and carries the uncompensated noise (see Fig. \ref{fig:exp_set-up}).

The 220~MHz beatnote that is detected and the 110~MHz correction signal that needs to be generated are outside the ADC and DAC baseband (between $\mathrm{0}$ and $f_\mathrm{{clock}}/2$).
For the ADC, one can perform an analog treatment of the signal: using a 200~MHz signal from a reference, the 220~MHz beatnote can be downconverted to 20~MHz (see Fig.~\ref{fig:exp_set-up}.a) so that it is sampled by the ADC. Alternatively, it has been shown~\cite{9599700} that the beatnote at $2 \times f_{\mathrm{AOM}}$ can be sent directly to the ADC and undersampled within the second Nyquist band at  $f'_{\mathrm{in}}=|f_{\mathrm{beatnote}} - 2f_{\mathrm{clock}}|=25.76~\mathrm{MHz}$ (see Fig.~\ref{fig:exp_set-up}.b).

\begin{figure}[h!]
    \centering
    \includegraphics[width=0.95\columnwidth]{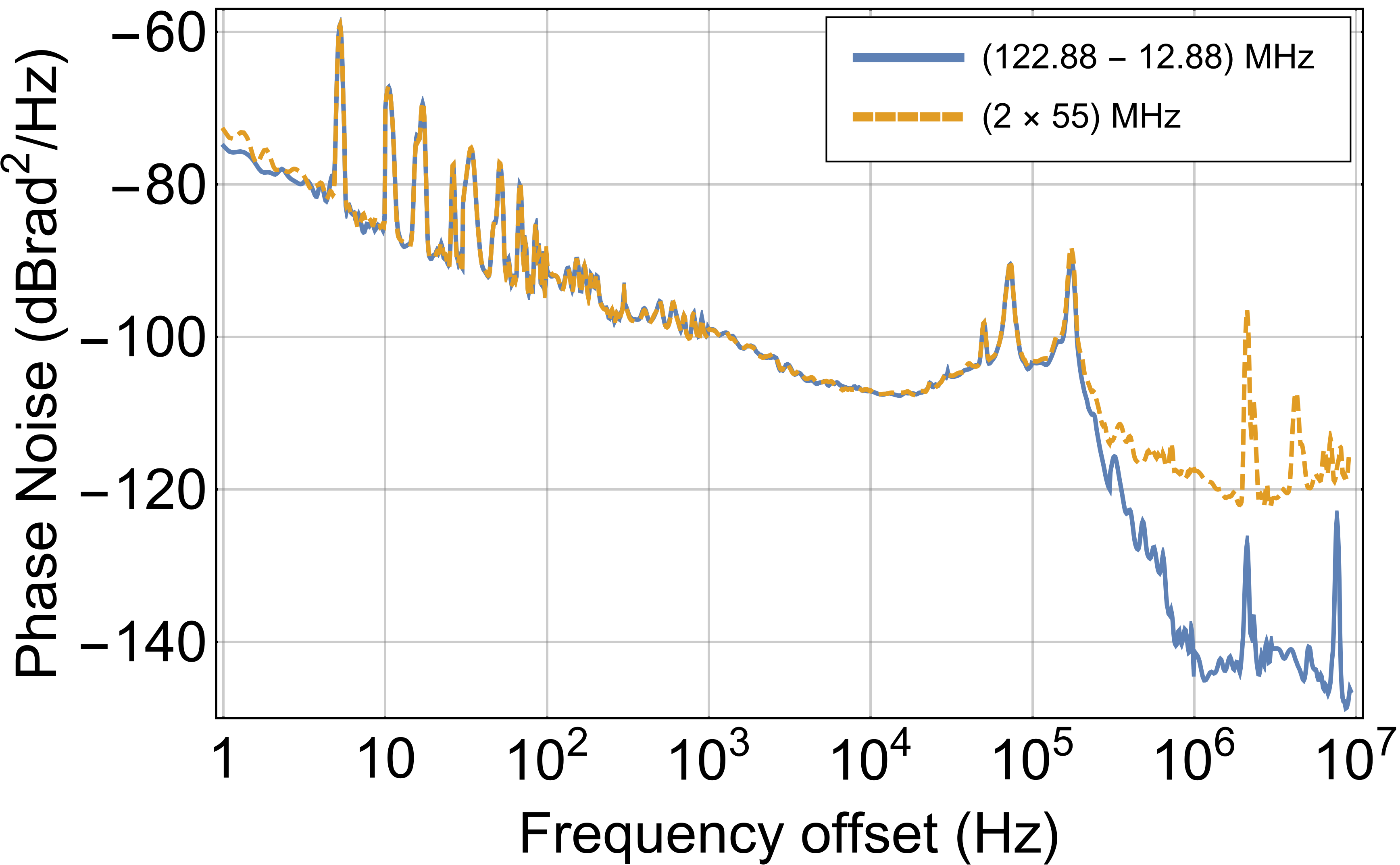}
    \caption{Phase-noise of the 110~MHz signal: (blue) obtained by selecting the aliased image of a 12.88~MHz signal or (yellow) obtained by frequency doubling a 55~MHz signal.}
    \label{fig:out110}
\end{figure}

For the DAC, one can similarly perform an analog treatment of the signal: a 55~MHz signal within the Red Pitaya baseband can be output and frequency doubled to drive the AOM at 110~MHz (see Fig.~\ref{fig:exp_set-up}.a). Here, we propose to instead output a signal at $f_{\mathrm{out}}=12.88~{\mathrm{MHz}}$, and to use the aliased image of the fundamental at $f_{\mathrm{clock}} - f_{\mathrm{out}} = {\mathrm{110~ MHz}} = f_{\mathrm{AOM}}$.
As the 3~dB bandwidth of the AOM is 30 MHz and the nearest spectral component at the output is only $2\times f_{\mathrm{out}} = 25.76$~MHz away, we use a  bandpass surface acoustic wave (SAW) filter\footnote{Golledge MP01327} to select the 110~MHz spectral component (see Fig.~\ref{fig:exp_set-up}.b). 
The filter was chosen despite its narrow bandwidth of 200~kHz for the very low thermal sensitivity of its group delay (measured to be below $\mathrm{10^{-7}~K^{-1}}$). It is not expected to impact the measurement in a laboratory environment, contrary to other larger-bandwidth existing SAW filters.
The phase-noise of the 110~MHz signals obtained either by frequency doubling a 55~MHz signal or by selecting the aliased image of a 12.88~MHz signal are indicated in Fig.~\ref{fig:out110}. Due to the sinc envelope of the DDS output, the amplitude of the signal at 110~MHz is a factor 6 smaller than at 55~MHz, but we see no degradation of the phase-noise by using the aliased output configuration. The lower sensitivity to environmental perturbations of digital electronics leads to an overall small improvement ($\lesssim 2$) of the link fractional frequency stability for integration times below 2000~s.  The lower phase-noise obtained for the aliased signal above $200~\mathrm{kHz}$ results from the narrow bandwidth of the SAW filter. 
We observe at $110~\mathrm{MHz}$ a larger influence of the $80~\mathrm{kHz}$ and $170~\mathrm{kHz}$ spurious signals already observed at $10~\mathrm{MHz}$.  Noise peaks in the 5-100~Hz range are caused by high current drawn by the RP16, inducing voltage noise on its power supply.

\begin{figure}[h!]
    \centering
    \includegraphics[width=0.95\columnwidth, clip]{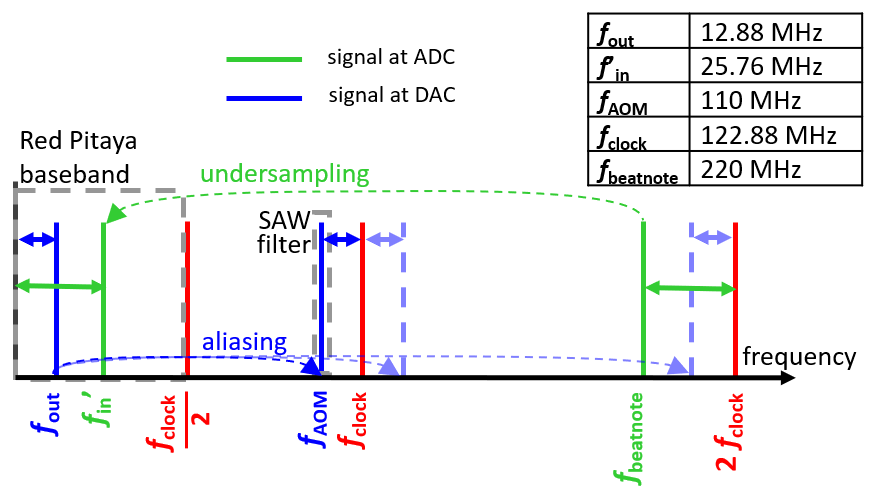}
    \caption{Digital ADC and DAC signal processing. ADC signal treatment relies on undersampling, where the signal with a frequency $f_{\mathrm{beatnote}} = 220$~MHz outside the Nyquist baseband is sampled as a signal inside the baseband with a frequency ${f'_\mathrm{in}=2f_{\mathrm{clock}}-f_{\mathrm{beatnote}}} = 25.76$~MHz. Regarding the ADC signal treatment, the DDS generates a signal at $f_\mathrm{{out}}=12.88$~MHz that carries the noise correction. It is aliased at $n{\times}f_{\mathrm{clock}} \pm f_{{out}}$, with $n$ an integer number. We use a SAW filter to select the $f_\mathrm{AOM} = f_\mathrm{clock} - f_\mathrm{out}=110$~MHz spectral component that drives the AOM.}
    \label{fig:ADC_DAC_signal}
\end{figure}

A summary of the frequency shifts we perform by aliasing at ADC and DAC is shown in Fig.~\ref{fig:ADC_DAC_signal}.

The beatnote detected by the ADC is numerically multiplied by a complex signal at $f_\mathrm{{demod}}$ (with $f_\mathrm{{demod}} = 20$~MHz for the analog ADC treatment, and $f_\mathrm{{demod}} = 25.76$~MHz for the digital ADC treatment). The phase information is obtained by extracting the real part of the product and filtering it by a finite impulse response (FIR) filter, realizing overall a cosine demodulation. It is then provided to a PI controller, the output of which drives a numerically controlled oscillator ($\mathrm{NCO_{out}}$)  at $f_\mathrm{out}$, that carries the noise correction.

The bandwidth of a PLL is ultimately limited by the delays of the system: for a given delay $\tau$, perturbations with frequencies higher than $1/2\tau$ can not be corrected. Therefore, we set the maximum bandwidth of the system to $1/4\tau$ (such that $2 \pi \tau f_\mathrm{pert} = \pi/2$, where $f_\mathrm{pert}$ is the perturbation frequency; this may still be close to the instability threshold, so a safer bandwidth of $1/8\tau$ is sometimes used instead as an upper limit in the literature).
The delays and resulting cumulative bandwidth of our system are indicated in Table~\ref{tbl:delaytable}.
It shows that the ADC and DAC conversions induce an irreducible delay of 125~ns, which sets a hard limit to 2~MHz for locks operated with Red Pitaya boards.
Other digital blocks (complex and real conversions, multiplications, adjustments of the number of bits) increase the delay by about $345~\mathrm{ns}$, while analog components play a dominant role, with 1.3~$\upmu$s delay due to the SAW filter and 2.5~$\upmu$s due to the AOM.
The propagation delay itself for a retro-reflected 90~m-long fiber further reduces the $1/4\tau$ bandwidth by about 17\%, down to 49~kHz. 
To emphasize the influence of each block, the phase shift it induces at 49~kHz is also indicated in Table~\ref{tbl:delaytable}.

For longer links, the propagation delay would play a more important role: for instance, a 500~m-long link would induce a delay of 2.5~$\upmu$s, making it the dominant contribution to the total delay. Longer links are also noisier, and would benefit from an arctan demodulation to correct large perturbations inducing phase flips. Arctan demodulation is more demanding in terms of digital processing and would further increase digital delays to 565~ns.

\begin{table}[h!]
\begin{tabular}{cc|c|c|c|}
\cline{3-5}
\multicolumn{1}{l}{} &
  \multicolumn{1}{l|}{} &
  \begin{tabular}[c]{@{}c@{}}Delay \\ $\tau$\end{tabular} &
  \begin{tabular}[c]{@{}c@{}}Cumulative \\ bandwidth\end{tabular} &
  \begin{tabular}[c]{@{}c@{}}Induced \\ dephasing \\ at 49 kHz\end{tabular} \\ \hline
\multicolumn{1}{|c|}{\multirow{2}{*}{\begin{tabular}[c]{@{}c@{}}Digital\\ electronics\end{tabular}}} &
  \begin{tabular}[c]{@{}c@{}}ADC and DAC\\ conversions\end{tabular} &
  125 ns &
  2~MHz &
  0.01 $\pi$ \\ \cline{2-5} 
\multicolumn{1}{|c|}{} &
  \begin{tabular}[c]{@{}c@{}}FIR and other\\ digital blocks\end{tabular} &
  345 ns &
  532~kHz &
  0.03 $\pi$ \\ \hline
\multicolumn{1}{|c|}{\multirow{2}{*}{\begin{tabular}[c]{@{}c@{}}Analog\\ electronics\end{tabular}}} &
  SAW &
  1.3 $\upmu$s &
  141~kHz &
  0.17 $\pi$ \\ \cline{2-5} 
\multicolumn{1}{|c|}{} &
  AOM &
  2.5 $\upmu$s &
  59~kHz &
  0.21 $\pi$ \\ \hline
\multicolumn{1}{|c|}{Fiber} &
  2 $~\times$ ~90~ m &
  880 ns &
  49~kHz &
  0.07 $\pi$ \\ \hline
\end{tabular}
\vspace{+0,6mm}
\caption{Delays, cumulative bandwidth and induced dephasing at 49~kHz due to different components of the system. For each component, the cumulative bandwidth is obtained from $1/4\Sigma \tau$, where the sum is performed on the delays of the component studied and of the components appearing above in the table.}
\label{tbl:delaytable}
\end{table}

\section{\label{auto_char} Automatic characterization}
To measure the noise rejection of an installed compensated fiber link, it is usually necessary that actually two links are set up in a reverse configuration, such that the output of the second link can be measured against the input of the first link. We propose here a direct digital noise-rejection evaluation method that does not require access to the remote end of the link.
It can be used on an existing system or efficiently deployed on multiple installations within an institute. The idea is to measure the rejection of sinusoidal perturbations for given proportional-integral (PI) controller parameters. 
We use it to experimentally find PI parameters that lead to the highest bandwidth (corresponding to the $1/4\tau$ bandwidth discussed previously) without increasing the noise at high frequencies and then divide the PI gains arbitrarily by a factor of two to avoid loop instabilities. 

\begin{figure}[h!]
    \centering
    \includegraphics[width=9.0cm, trim = {0 1.1cm 0 0.8cm}, clip]{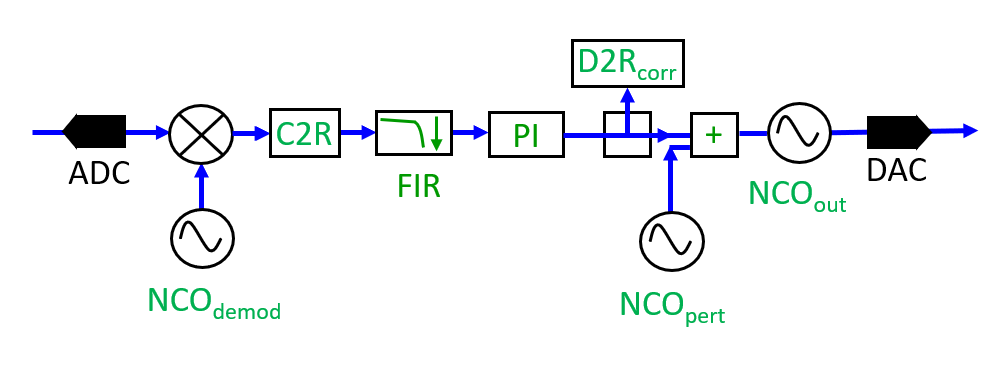}
    \caption{Red Pitaya design adapted for disturbance rejection measurements: $\mathrm{NCO_{pert}}$ adds a perturbation to the correction signal, and the cancellation of this perturbation is measured by $\mathrm{D2R{corr}}$. $\mathrm{D2R}$: data to ram acquisition, $\mathrm{NCO}$: numerically controlled oscillator. $\mathrm{FIR}$: finite impulse response filter. $\mathrm{C2R}$: complex to real conversion.} 
    \label{fig:double_iq_pid_vco_charac}
\end{figure}

For this purpose, we induce a frequency modulation at $f_\mathrm{pert}$ of the output signal: a sinusoidal perturbation at $f_\mathrm{pert}$ is numerically added to the correction signal that drives $\mathrm{NCO_{out}}$. 
This simulates the effect of a frequency perturbation arising inside the fiber link. 
The disturbance rejection by the PLL is numerically measured right after the PI block.
A scheme of the modified Red Pitaya architecture is indicated in Fig~\ref{fig:double_iq_pid_vco_charac} and available at~\cite{oscimpPLL}.

\begin{figure}[h!]
    \centering
    \includegraphics[width=9.5cm]{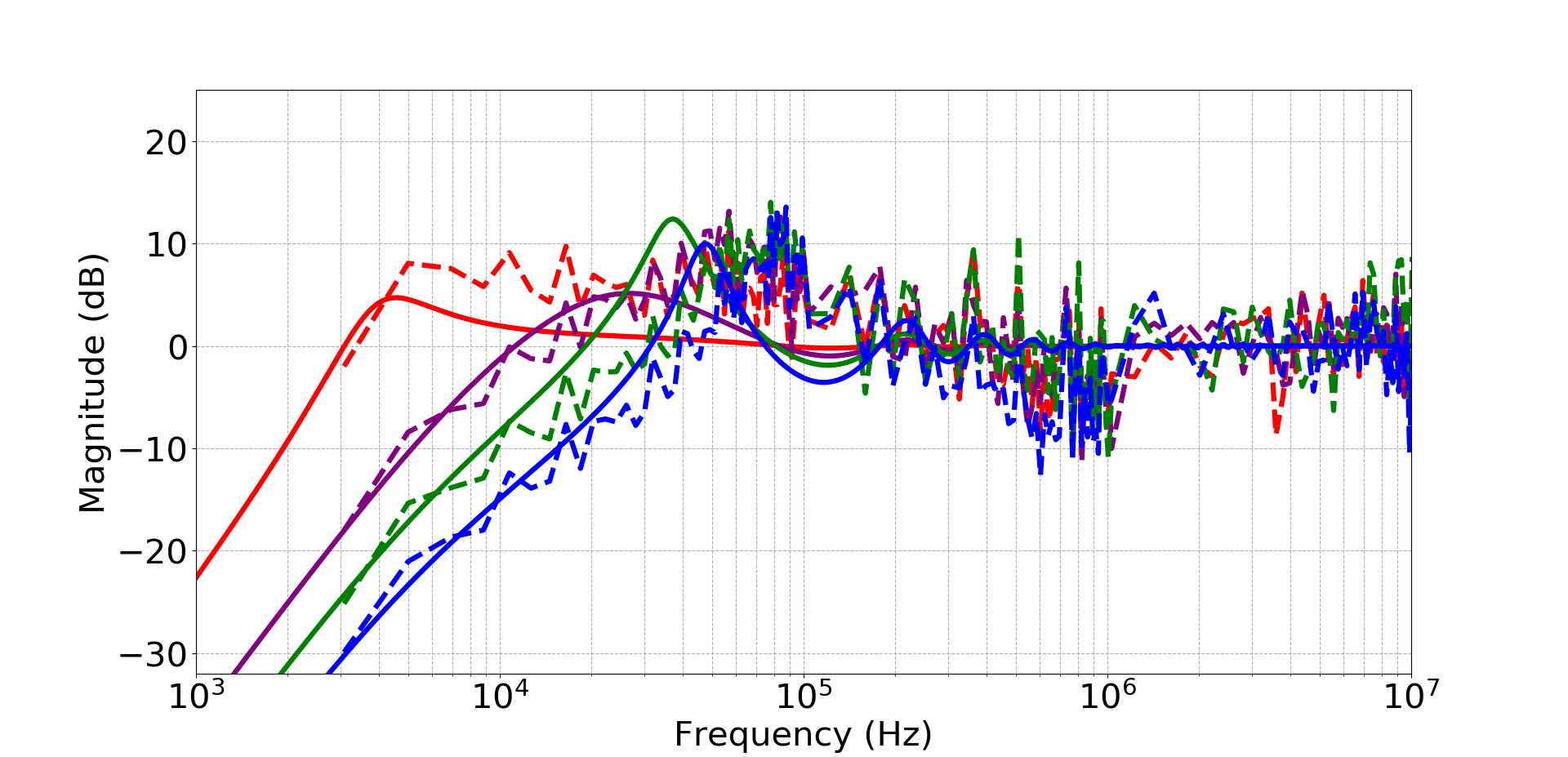}
    \caption{Disturbance rejection of the system with a 90~m-long fiber link. Dashed lines correspond to the experimental disturbance rejection. Solid lines correspond to the modeled disturbance rejection. Red curves correspond to the lowest PI gains; purple curves correspond to PI gains 5 times larger, green curves to PI gains 10 times larger and blue curves to PI gains 20 times larger than the ones of the red curve.
   }
    \label{fig:transfer_link}
\end{figure}

$f_\mathrm{pert}$ is swept between 3~kHz and 10~MHz for both open loop configuration (no PI gain) and closed loop configuration, with adjusted PI gains. The disturbance rejection (ratio of the perturbation amplitudes in the closed and open loop configurations) measured with a 90~m-long link is shown in Fig.~\ref{fig:transfer_link} with dashed lines for four sets of PI gains. It is modelled with solid lines using  optical system transfer functions that were first described in works of \cite{jiang:2010, bercy:2015}. We chose an equivalent approach in the discrete time domain and used  $z$-transforms formalism that takes into account analog, digital and optical propagation delays, with the global gain as the only free parameter. In the case of cosine demodulator the mixer conversion gain (rad/V) is dependent on the magnitude of the input signal. Thus it creates a scaling of the total loop gain, which is then dependent on the photodiode beatnote amplitude.
The model does not take into account the  spurious signals observed around 80 and 170~kHz. 
At low frequencies there is a good agreement between the model and experimental disturbance rejection curves. 
For perturbation frequencies above 200~kHz, the SAW filter acts as a very efficient low-pass filter, regardless of the PI gains. As it is then the ratio of two small quantities, the experimental disturbance rejection curve becomes noisy, but its value consistently averages to 0~dB. 
For all experimental curves, we observe a shift of the overshoot towards 80~kHz with respect to the model. We attribute this to the proximity of the 80~kHz instability mentioned previously (see Fig. \ref{fig:out110}). 
The highest bandwidth that can be reached with this system is about 40~kHz (see blue curve Fig.~\ref{fig:transfer_link}), consistent with Table~\ref{tbl:delaytable}. This configuration is close to the instability threshold (where oscillations induce excess noise above 200~kHz), so we rather favored the global system stability and reduced these PI gains by a factor two, leading to a 3~dB bandwidth around 30~kHz (see green curve Fig.~\ref{fig:transfer_link}).

\begin{figure}[h!]
    \centering
    \includegraphics[width=9.5cm]{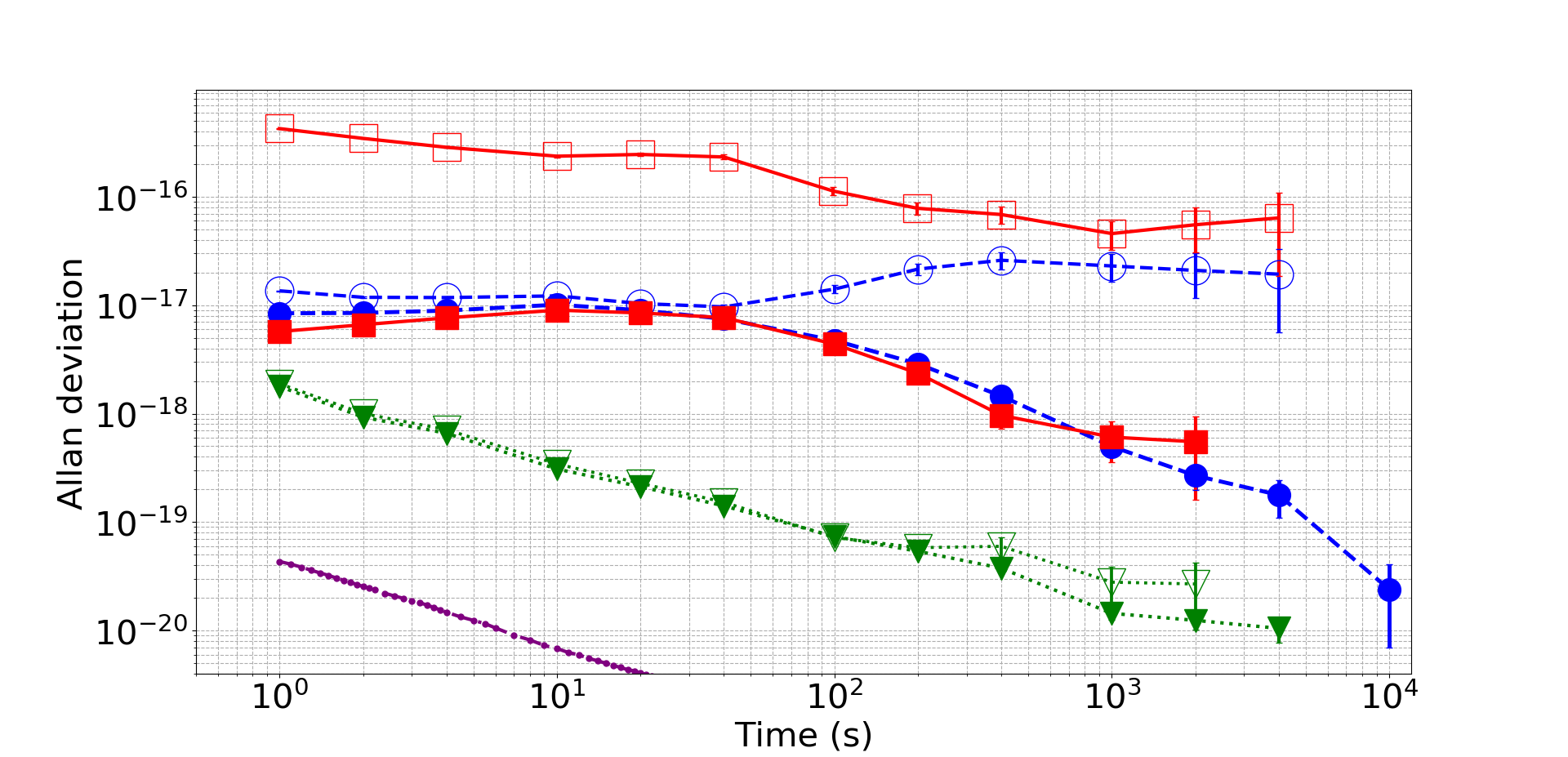}
    \caption{Fractional frequency instabilities of the optical signal at PD2 for different experimental set-ups. The instability is measured on the 110~MHz beatnote and scaled to the optical frequency of 194~THz with a leverage factor of $5.7 \times \mathrm{10^{-7}}$. 
    Open symbols: free running system. Full symbols: closed loop measurement. Green triangles: no physical system (see text). Blue circles: with 1~m link. Red squares: with 90~m link. Purple line: Lower limit set by the hydrogen maser, given by the maser fractional frequency instability multiplied by the leverage factor. Lines are a guide to the eye.  All measurements were taken without any specific shielding of the system. 
    }
    \label{fig:adev}
\end{figure}

We used this method to choose PI gains and then measured the resulting signal frequency instability of the monitoring 110~MHz signal (see Fig.~\ref{fig:exp_set-up}) for both open and closed loop in three situations: first, with a 1~m-long fiber link, then with a 90~m-long fiber link, and last, for reference, without the physical system constituted of the AOM and fibers.
Results are indicated in Fig.~\ref{fig:adev} by open symbols for the open loop and full symbols for the closed loop.

For the 1~m-long and the 90~m-long links (blue circles in Fig.~\ref{fig:adev} and red squares in Fig.~\ref{fig:adev}, respectively), the plateau in stability for the 1 - 40~s range points towards flicker frequency noise, consistent with the $1/f$ slope we observe in their power spectrum for frequencies higher than $10^{-2}$~Hz (not shown).

For the 1~m-long link, the open and closed loop curves overlap for the first 40~s of integration, indicating that for short integration times the stability of the system is limited by the reduction of the lock bandwidth due to the optical components (AOM, couplers, Faraday mirrors, photodiodes).
For closed loop configuration, the fractional frequency instability is below $10^{-17}$ for all integration times and reaches $1\times 10^{-18}$ for integration times longer than 500~s.

For the 90~m-long link, the open loop signal is degraded to the $ 10^{-16}$ range, illustrating the reduction of the signal stability due to the 90~m fiber. When the loop is closed, the signal shows the same instability as that of the 1~m-long fiber link, showing that we reached the limit of the system. With an instability at most in the $\mathrm{10^{-18}}$ range for all integration times and that averages down to  $6\times 10^{-19}$ for 2000~s integration time, the set-up fully meets the requirements for frequency transfer within an institute. 

For reference, we removed the physical system containing the AOM and the optical fibers. The 110~MHz component selected by the SAW filter is split in two. One part is directly sampled by the ADC of RP16 (with $\mathrm{NCO_{demod}}$ set to 12.88~MHz). The other part is sent to a frequency counter to measure the in-loop fractional frequency instability of the 110~MHz signal. It is then scaled to optical frequencies and shown with green triangles in Fig.~\ref{fig:adev}. The open and closed loop curves overlap for all integration times, indicating the limitations due to the RP16 board.

\section{Conclusion}

We compared RP16 and RP14 boards for Doppler cancellation of fiber links using the fully open-source FPGA framework OscimpDigital\cite{oscimp}. With its design explicitly targeting radio frequency instrumentation, RP16 proved more adapted for metrology applications. Here, we focused on the development of a local fiber network for ultra-stable signal distribution.

The noise arising from the fiber links is compensated by a digital PLL hosted on the RP16 board with a novel signal processing based on aliasing at the RP16 DAC and undersampling at the RP16 ADC.
This alleviated the need for many environment-sensitive analog components and resulted in a compact set-up easy to duplicate. 
We have measured the delays of the system and estimated the resulting PLL bandwidth. The delay due to ADC and DAC conversions set an upper limit of the bandwidth for any Red Pitaya-based system at 2~MHz. Taking all the delays of the system into account brings the upper limitation of the PLL bandwidth to 49~kHz. For a 90~m-long link, propagation delay contributes to about 17\% of the total delay, therefore the total bandwidth is significantly reduced for longer links.

Lastly, we have developed an automatic measurement of the disturbance rejection for given PI gains of the PLL, therefore providing a rigorous method to measure the bandwidth. The maximum bandwidth we measured was in agreement with the calculated one, but the system was close to the instability threshold. For a 30\% lower bandwidth, the system was stable and we measured fractional frequency instabilities in the $10^{-18}$ range for all timescales. 
Performances could in principle be improved by the use of a second AOM at the remote end of the system to discriminate unwanted reflections~\cite{Amy-Klein2022_reflections}, but the current system is already perfectly suitable for local ultra-stable frequency distribution within a metrology institute. 

\section*{Acknowledgments}
This work was supported by Agence Nationale de la Recherche (ANR-21-CE47-0006-01 CONSULA, ANR-10-LABX-48-01 First-TF, and ANR-11-EQPX-0033 Oscillator-IMP), the Région Bourgogne Franche-Comté and the EIPHI Graduate School (contract ANR-17-EURE-0002). 
The authors would like to thank Gonzalo Cabodevila for discussions on PLL characterization, Vincent Giordano for fruitful discussions and Claudio Calosso for careful reading of the manuscript.

\nocite{*}
\bibliography{biblio}

\end{document}